\documentclass[twocolumn, onecolappendix]{aastex63}

\usepackage{amsmath}
\usepackage{mathrsfs}
\usepackage{afterpage}
\usepackage{scalerel}
\usepackage{natbib}
\bibpunct[; ]{(}{)}{;}{a}{}{,}
\usepackage{multirow}
\usepackage{graphics}
\usepackage{threeparttable}
\usepackage[varg]{txfonts}
\usepackage{xcolor}
\usepackage{bm}

\shorttitle{The Most Obscured AGNs in the XMM-SERVS Fields}
\shortauthors{Wei Yan et al.}

\newcommand{\servs}{{XMM-SERVS}}

\newcommand{\cdens}{cm$^{-2}$}
\newcommand{\nh}{$N_{\rm H}$}
\newcommand{\fct}{$f_{\rm CT}$}

\begin{document}

\title{The Most Obscured AGNs in the XMM-SERVS Fields}

\author[0000-0001-9519-1812]{Wei Yan}
\email{wmy5111@psu.edu}
\affiliation{Department of Astronomy and Astrophysics, 525 Davey Lab, The Pennsylvania State University, University Park, PA 16802, USA}
\affiliation{Institute for Gravitation and the Cosmos, The Pennsylvania State University, University Park, PA 16802, USA}

\author[0000-0002-0167-2453]{W. N. Brandt}
\affiliation{Department of Astronomy and Astrophysics, 525 Davey Lab, The Pennsylvania State University, University Park, PA 16802, USA}
\affiliation{Institute for Gravitation and the Cosmos, The Pennsylvania State University, University Park, PA 16802, USA}
\affiliation{Department of Physics, 104 Davey Lab, The Pennsylvania State University, University Park, PA 16802, USA }

\author[0000-0002-4436-6923]{Fan Zou}
\affiliation{Department of Astronomy and Astrophysics, 525 Davey Lab, The Pennsylvania State University, University Park, PA 16802, USA}
\affiliation{Institute for Gravitation and the Cosmos, The Pennsylvania State University, University Park, PA 16802, USA}

\author[0000-0002-1653-4969]{Shifu Zhu}
\affiliation{Department of Astronomy and Astrophysics, 525 Davey Lab, The Pennsylvania State University, University Park, PA 16802, USA}
\affiliation{Institute for Gravitation and the Cosmos, The Pennsylvania State University, University Park, PA 16802, USA}

\author[0000-0002-4945-5079]{Chien-Ting J. Chen}
\affiliation{Science and Technology Institute, Universities Space Research Association, Huntsville, AL 35805, USA}
\affiliation{Astrophysics Office, NASA Marshall Space Flight Center, ST12, Huntsville, AL 35812, USA}

\author[0000-0003-1468-9526]{Ryan C. Hickox}
\affiliation{Department of Physics and Astronomy, Dartmouth College, Hanover, NH 03755, USA} 
 
\author[0000-0002-9036-0063]{Bin Luo}
\affiliation{School of Astronomy and Space Science, Nanjing University, Nanjing, Jiangsu 210093, China}
\affiliation{Key Laboratory of Modern Astronomy and Astrophysics (Nanjing University), Ministry of Education, Nanjing 210093, China}

\author[0000-0002-8577-2717]{Qingling Ni}
\affiliation{Max-Planck-Institut f\"{u}r extraterrestrische Physik (MPE), Gie{\ss}enbachstra{\ss}e 1, D-85748 Garching bei M\"unchen, Germany}

\author[0000-0002-5896-6313]{David M. Alexander}
\affiliation{Centre for Extragalactic Astronomy, Department of Physics, Durham University, South Road, Durham, DH1 3LE, UK}

\author[0000-0002-8686-8737]{Franz E. Bauer} 
\affiliation{Instituto de Astrof{\'{\i}}sica, Facultad de F{\'{i}}sica, Pontificia Universidad Cat{\'{o}}lica de Chile, Campus San Joaquín, Av. Vicuña Mackenna 4860, Macul Santiago, Chile, 7820436} 
\affiliation{Centro de Astroingenier{\'{\i}}a, Facultad de F{\'{i}}sica, Pontificia Universidad Cat{\'{o}}lica de Chile, Campus San Joaquín, Av. Vicuña Mackenna 4860, Macul Santiago, Chile, 7820436} 
\affiliation{Millennium Institute of Astrophysics, Nuncio Monse{\~{n}}or S{\'{o}}tero Sanz 100, Of 104, Providencia, Santiago, Chile}

\author[0000-0002-8853-9611]{Cristian Vignali}
\affiliation{Dipartimento di Fisica e Astronomia ``Augusto Righi'', Alma Mater Studiorum, Universit\`a degli Studi di Bologna, Via Gobetti 93/2, 40129 Bologna, Italy}
\affiliation{INAF -- Osservatorio di Astrofisica e Scienza dello Spazio di Bologna, Via Gobetti 93/3, I-40129 Bologna, Italy}

\author[0000-0003-0680-9305]{Fabio Vito}
\affiliation{INAF -- Osservatorio di Astrofisica e Scienza dello Spazio di Bologna, Via Gobetti 93/3, I-40129 Bologna, Italy}

\accepted{April 2023}

\begin{abstract}

We perform X-ray spectral analyses to derive characteristics (e.g., column density, X-ray luminosity) of $\approx$10,200 active galactic nuclei (AGNs) in the XMM-Spitzer Extragalactic Representative Volume Survey (XMM-SERVS), which was designed to investigate the growth of supermassive black holes across a wide dynamic range of cosmic environments. Using physical torus models (e.g., \textit{Borus02}) and a Bayesian approach, we uncover 22 representative Compton-thick (CT; \nh  $\;>\; 1.5\times10^{24}\; \rm cm^{-2}$) AGN candidates with good signal-to-noise ratios as well as a large sample of 136 heavily obscured AGNs. We also find an increasing CT fraction (\fct ) from low ($z<0.75$) to high ($z>0.75$) redshift. Our CT candidates tend to show hard X-ray spectral shapes and dust extinction in their SED fits, which may shed light on the connection between AGN obscuration and host-galaxy evolution.

\end{abstract}


\section{Introduction}

Luminous unobscured (type 1) active galactic nuclei (AGNs) have been well-studied ever since they were discovered over 50 years ago. Thanks to their generally high luminosities, unobscured AGNs often dominate over the host-galaxy light at most wavelengths, making them relatively easy to observe and study. However, it is now known that half or more of AGNs are obscured by gas and dust (e.g., \citealt{hick07abs, bran15, mate17obsc}). The existence of many obscured (type 2) AGNs has direct implications for the growth history of supermassive black holes (SMBH) in galactic centers across cosmic time (e.g., \citealt{bran10, hick18}) as well as the origin of the cosmic X-ray background (CXB; e.g., \citealt{gill07cxb, trei09, ueda14cxb, aird15nustar}). 

Some recent progress has suggested the existence of a large population of Compton-thick (CT) AGNs with intrinsic column densities of $N_{\rm H}$ $ > 1.5 \times 10^{24}\; \rm cm^{-2}$ (e.g., \citealt{lans15nustarqso, lalo23}). CT sources comprise a large fraction of lower-luminosity AGNs (e.g., \citealt{goul12comp, ricc17nustarwise}), which contribute to models of the CXB spectrum (e.g., \citealt{ueda14cxb}). Moreover, for less-powerful AGNs (i.e., Seyfert galaxies), the classic `unified model' is largely successful in explaining obscuration by varying viewing angles of an obscuring `torus' (e.g., \citealt{netz15unified}). However, it remains unclear if this picture also holds for powerful AGNs. The observed dependence of AGN obscuration upon luminosity indicates a departure from the simplest unified model. The most-powerful AGNs may also be obscured by starbursts (e.g., \citealt{ball08agnsb}) or larger-scale gas clouds driven to the center of the galaxy by violent mergers or instabilities (e.g., \citealt{dima05qso, hopk08disk_aph}). Different AGN fueling mechanisms produce different distributions of \nh\ and CT fractions (\fct ; e.g., \citealt{drap12}).

\begin{figure*}[t]
\epsscale{1.1}
\plotone{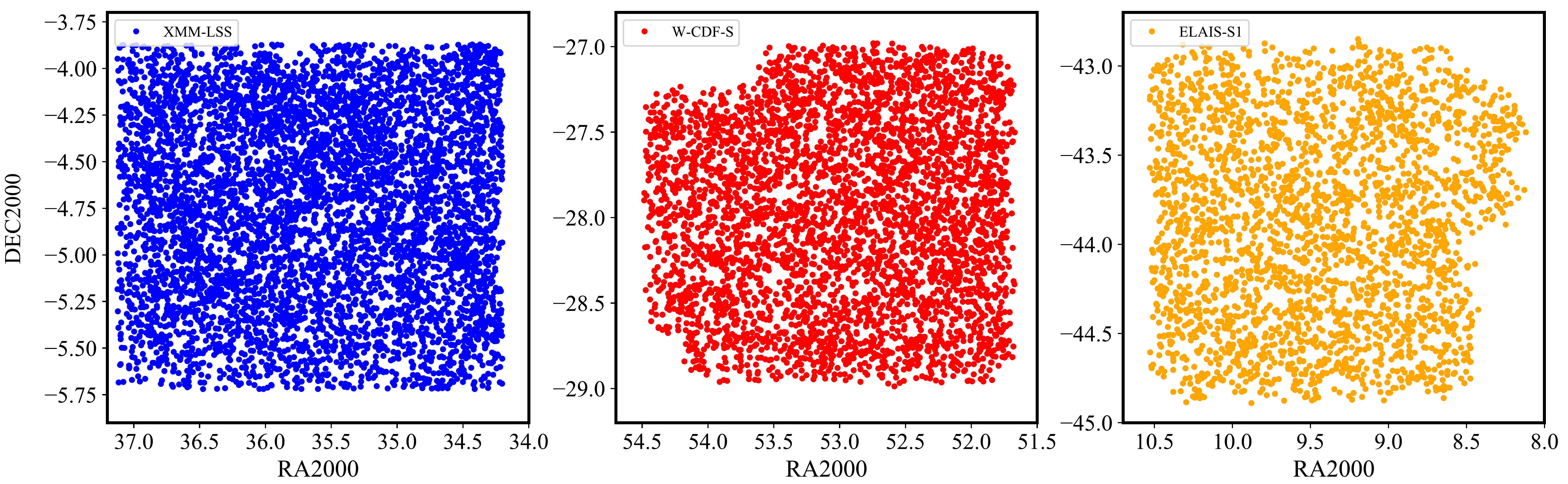}
\caption{Sky maps of the three fields covered by \servs . Blue, red, and orange points are the X-ray point sources detected in the XMM-LSS, W-CDF-S, and ELAIS-S1 fields, respectively. \label{fig:sky}}
\end{figure*}

While \fct\ for more luminous AGNs has implications for their contribution to the CXB and galaxy evolution, relatively few luminous CT AGNs with intrinsic column densities above $1.5\times10^{24}\; \rm cm^{-2}$ have been directly confirmed (e.g., \citealt{gill11ct,iwas12, lans17nustar, vito18ho, yan19nustar}). The obscuration makes the CT AGNs difficult to find and study. For example, the photoelectric absorption cut-off (e.g., at observed $\approx\; 10$ keV for a $z = 0.2$ AGN absorbed by a column density of $\approx\;10^{24}\; \rm cm^{-2}$) dramatically reduces the X-ray flux. This may bias the measured spectral parameters, such as the intrinsic power-law photon index $\Gamma$ or \nh\ if fitting spectra with low counts or a limited energy range. Also, CT levels of absorption deeply suppress the primary continuum, revealing strong $\rm Fe\; K\alpha$ line emission at 6.4--7 keV and a Compton-reflection `hump' at $\sim$ 20--30 keV. Due to observational limitations, we have previously obtained only weak constraints on the distribution of the obscuring column density \nh\ of luminous AGNs. 

Thanks to recent large, sensitive X-ray surveys, we are able to detect large representative samples of the AGN population. For example, the XMM-Spitzer Extragalactic Representative Volume Survey (\servs\ ; \citealt{chen18lss, ni21servs}) provides over 10,200 representative AGNs from its wide XMM-Newton \mbox{X-ray} coverage of three well-studied fields: XMM-Large Scale Structure (XMM-LSS; 5.3 ${\rm deg}^2$), Wide Chandra Deep Field-South (W-CDF-S; 4.6 ${\rm deg}^2$), and European Large-Area ISO Survey-S1 (ELAIS-S1; 3.2 ${\rm deg}^2$). Figure~\ref{fig:sky} shows the sky maps of these three fields. The large \servs\ AGN population also generally has well-established redshifts and spectral energy distribution (SED) fits (e.g., \citealt{zou22sed}), derived from multi-wavelength spectroscopic and photometric surveys covering from the X-ray to far-infrared (far-IR). Furthermore, all three \servs\ fields are Legacy Survey of Space and Time (LSST) deep-drilling fields. Therefore, this substantial number of AGNs in these prime survey fields also provides a representative AGN sample for many future studies, characterizing the AGN population and its distinguishable features (e.g., obscuration, variability, and host properties) for the coming decades.

In this work, we systematically extract X-ray spectra and perform spectral analyses to derive AGN characteristics (e.g., \nh , photon index, X-ray luminosity) and select CT AGNs in \servs . Using a physical torus model and a Bayesian analysis approach, we uncover 22 representative CT AGN candidates with good signal-to-noise ratios (SNRs) as well as a large sample of the heavily-obscured (HO) AGNs (\nh\ $> \;5\times10^{23}\; \rm cm^{-2}$). The paper is organized as follows: Section 2 details the X-ray data analyses, and multi-wavelength SED analyses are discussed in Section 3. We discuss our results in Section 4 and then summarize the paper in Section 5. Throughout the paper, we assume a $\Lambda$CDM cosmology with $H_0=69 \rm {km\;s}^{-1}{\rm Mpc}^{-1}$, ${\rm \Omega}_{\rm M}=0.286$, and ${\rm \Omega}_{\rm \Lambda}=0.714$ \citep{wrig06coscal}.

\section{CT sample selection}

\begin{figure*}[t]
\epsscale{1.1}
\plotone{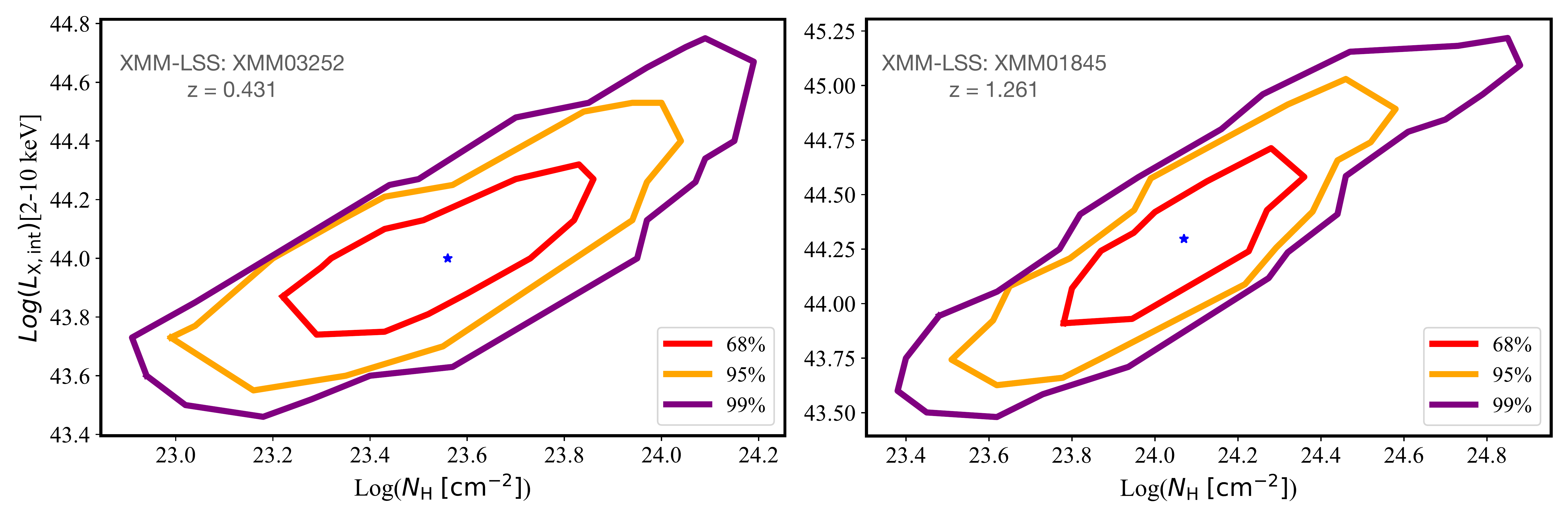}
\caption{Examples of the Bayesian spectral analysis results for our heavily-obscured sample (\textit{left}) and CT candidate sample (\textit{right}) in \servs . Each figure shows the posterior probability distribution function in the two-dimensional space of intrinsic X-ray luminosity in the 2--10 keV band and \nh\ from the X-ray spectral analyses. The star in the center represents the median luminosity and column density. The red, orange, and purple lines mark 68\%, 95\%, and 99\% contours of the posterior probability distribution.  \label{fig:example1}}
\end{figure*}

\subsection{X-ray spectral analyses with BXA}

We uniformly extract the X-ray spectra of all X-ray point sources in the three \servs\ fields (Figure~\ref{fig:sky}) with SAS tasks. First, we use EREGIONANALYSE to obtain an optimum radius for the source region, defined as a circle centered around the individual source with a radius automatically chosen to maximize the SNR at \mbox{2--10} keV. Then, we use MULTIESPECGET to extract source spectra from MOS1, MOS2, and pn, respectively. We conduct a similar procedure to obtain the individual background spectrum from a background region, defined as an annulus centered around the source region with an outer radius of 90 arcsec. For each individual source, we also exclude the overlap of the source regions of other sources from the selected source and background regions. We then generate the corresponding redistribution matrix and ancillary calibration files. Finally, for sources detected in more than one detector, after ensuring the spectral files from each detector of each source have a common spectral range, we use EPICSPECCOMBINE to co-add the spectra from all detectors as a single combined spectrum for spectral analysis, in order to further increase the SNR.

We then perform Bayesian spectral analyses on all extracted spectra uniformly and systematically without making a prior selection. We adopt the Bayesian X-ray Analysis package (BXA; \citealt{buch14baye, buch15}) to fit the X-ray spectra of all sources in \servs\ with \textit{Borus02}, a physical torus model, to estimate the suppression of AGN X-ray spectra by the obscuring material. \textit{Borus} \citep{balo18borus} is based on radiative-transfer calculations with an approximately toroidal geometry of neutral gas, appropriate for CCD-resolution X-ray spectra. This work uses \textit{Borus02}, which constructs a torus geometry of a smooth spherical distribution of neutral gas, with conical cavities along the polar direction. We adopt the following model in \textit{XSPEC} \citep{arna96}: $a*phabs*(Borus02 + zphabs*cabs*cutoffpl + b*cutoffpl)$. In this model, \textit{Borus02} contains the spectral components arising from reprocessing in the torus; $phabs$ accounts for foreground Galactic absorption; $zphabs*cabs$ represents line-of-sight intrinsic absorption at the redshift of the X-ray source, including Compton-scattering losses out of the line of sight; $cutoffpl$ represents the intrinsic continuum in the 2--10 keV band; and two constants $a$ and $b$ stand for normalization and the leaked or scattered unabsorbed reflection of the intrinsic continuum, respectively. We find that the number of sources with extreme obscuration is not strongly affected by linking or not linking the relevant column densities. Therefore, to avoid possible degeneracies, we decrease the number of free parameters by linking column densities in the spectral modeling. The model includes five free parameters: the covering factor (related to the half-opening angle of the torus), the line-of-sight column density $\log N_{\rm H}$, the slope of the intrinsic power-law spectrum $\Gamma$, and the two normalization parameters. We adopt a Gaussian prior for the spectral index $\Gamma$ with a mean of 1.8 and standard deviation of 0.3 (e.g., \citealt{nand94xspec}), and a flat prior for the line-of-sight obscuration $\log N_{\rm H}$ in the range of 20--25 (e.g., \citealt{geor17xspec}). The Hamiltonian Monte Carlo code Stan is used for Bayesian statistical inference. The source redshift is always fixed as its spectroscopic redshift when available. Otherwise, we allow the photometric redshift (photo-z) to vary between its upper and lower limits with a 68\% significance level derived based on forced photometry (e.g., \citealt{zou22sed}). The quality of the photometric redshift can be further examined using the probability associated with the peak redshift ($peak-prob$) and the photo-z quality indicator $Q_z$ from the XMM-SERVS catalogs (\citealt{chen18lss, zou21}).

Furthermore, to properly account for cases in which the \nh\ probability distribution function (PDF) has multiple peaks due to large uncertainties from limited counts, we examine the output PDFs of all spectra of individual \servs\ sources generated by our Bayesian analysis, in order to select representative reliable HO samples as well as CT candidates as explained below.

\subsection{Sample selection with PDFs}

\begin{figure*}[t]
\epsscale{1.1}
\plotone{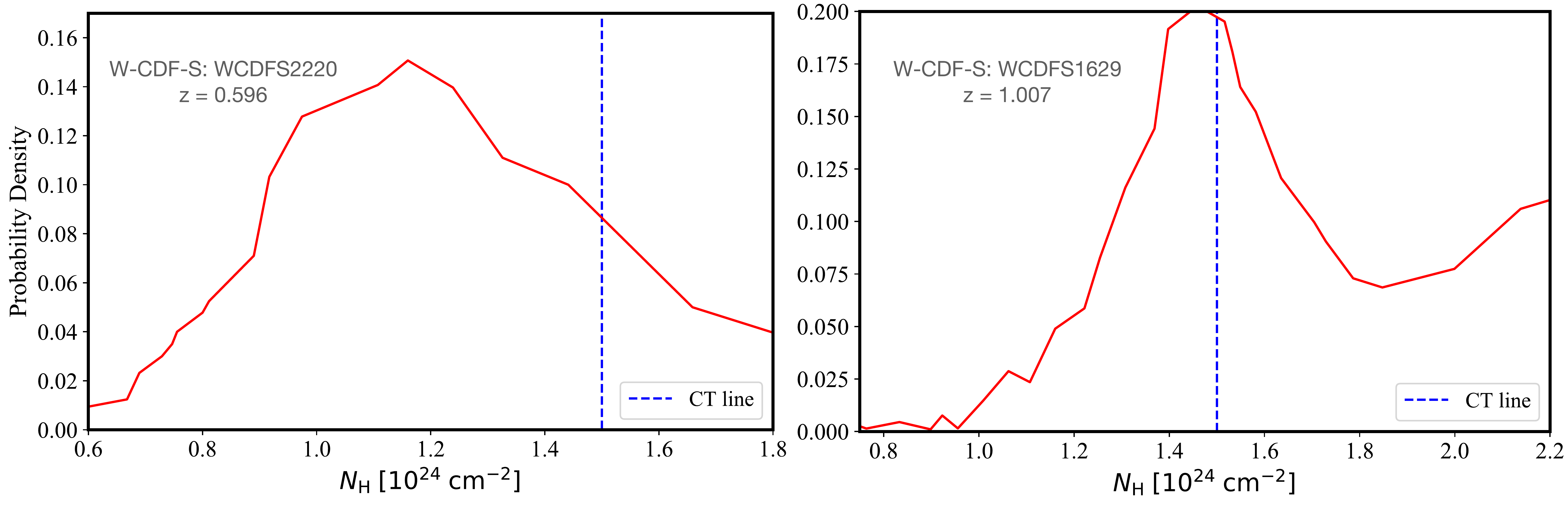}
\caption{Examples of the \nh\ PDFs (red lines) for our heavily-obscured sample (\textit{left}) and CT candidate sample (\textit{right}) in \servs\ obtained from the Bayesian analysis. The blue line marks the CT threshold at $N_{\rm H} = 1.5\times10^{24}\; \rm cm^{-2}$ .  \label{fig:example2}}
\end{figure*}

\begin{figure}
\includegraphics[scale=0.15]{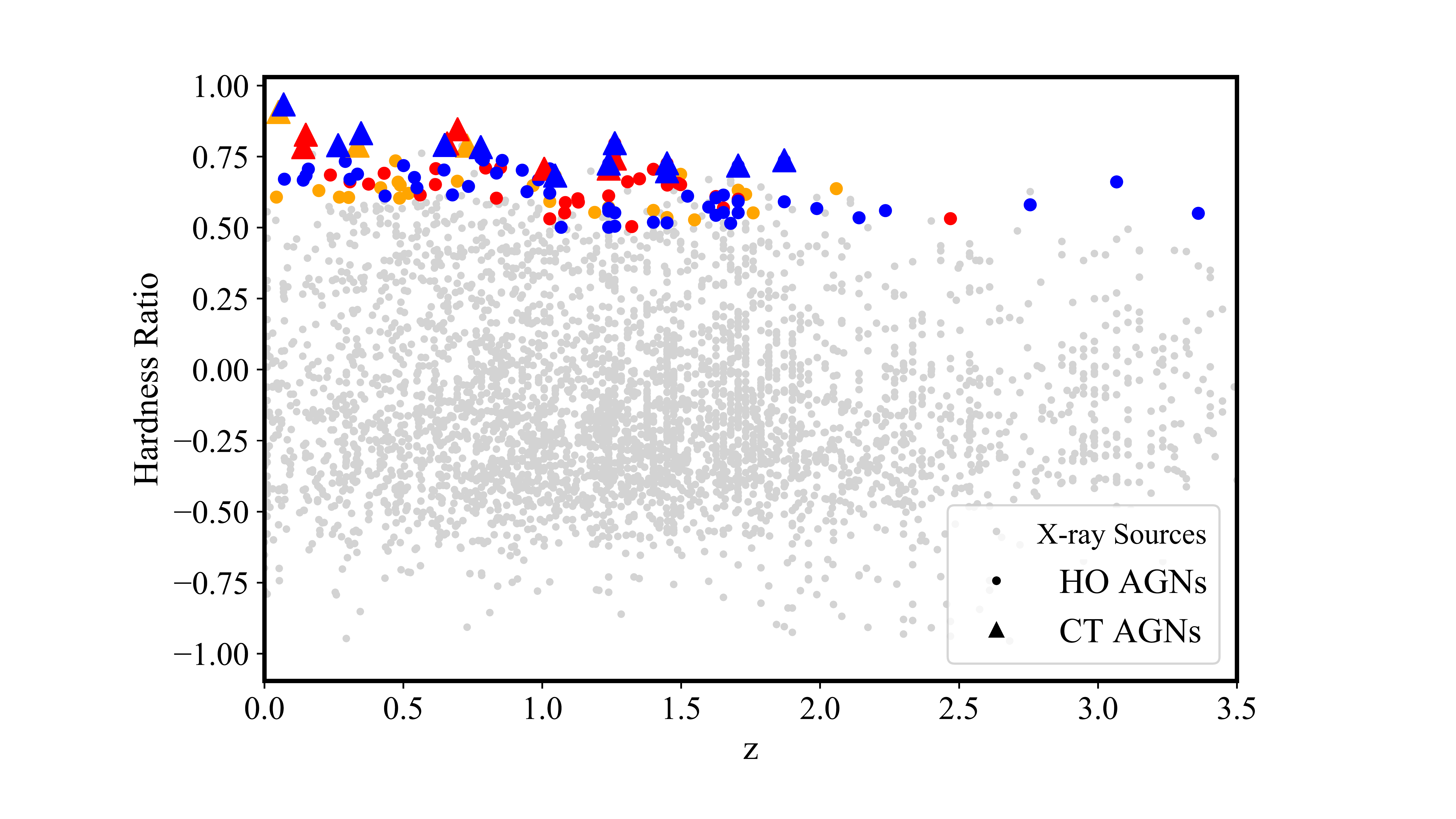}
\caption{The distribution of hardness ratio versus redshift of the selected heavily-obscured (\textit{circles}) and CT (\textit{triangles}) AGNs. The hardness ratio is adopted from XMM-SERVS catalogs, defined as $(H-S)/(H+S)$, where \textit{H} is the total net counts divided by the total exposure time in the hard band (2--10 keV), and \textit{S} is the total net counts divided by the total exposure time in the soft band (0.5--2 keV). Triangles represent CT candidates, and circles represent the heavily-obscured sources in all three fields. Blue, red, and orange colors mark the selected CT candidates in the XMM-LSS, W-CDF-S, and ELAIS-S1 fields, respectively (22 in total). We also show all the detected X-ray sources in the XMM-LSS field (grey dots) for comparison. The CT candidates, as expected, tend to be among the hardest sources in the three fields. \label{fig:hr}}
\end{figure}

We define a source as a CT candidate when its posterior \nh\ probability over the CT threshold (\nh\ $= \;1.5\times10^{24}\; \rm cm^{-2}$; $P_{\rm CT}$) is above 50\%. As a result, we uncover a total of 22 CT candidates (12 in XMM-LSS, 7 in W-CDF-S, and 3 in ELAIS-S1). We also select a large number of HO AGNs in the three fields (59 in XMM-LSS, 43 in W-CDF-S, and 35 in ELAIS-S1), where the posterior \nh\ probability over $ \;5\times10^{23}\; \rm cm^{-2}$ ($P_{HO}$) is above 50\%. The numbers of selected objects are summarized in Table 1. For our selected CT and HO samples, 8 (out of 22) CT candidates and 44 (out of 136) HO AGNs have spectroscopic redshift. We also list a few examples of the HO AGNs in Table 2 and the CT candidates in Table 3. We show some examples of the Bayesian spectral analysis results for these selected sources in Figure~\ref{fig:example1}, and the \nh\ PDF in Figure~\ref{fig:example2} with the CT threshold marked as blue dashed lines. In particular, the CT candidates are among the hardest sources in the parent sample (Figure~\ref{fig:hr}). These candidates also have good SNR in the hard X-ray band.

Based on the PDFs of the 22 CT candidates, we notice that none of these sources has a 100\% probability of being CT. Here we calculate the sum of the probability above the CT threshold for each source. By summing only the fraction of the PDF of each source above the CT threshold, we derive the total expected number of CT sources in \servs\ as $\approx$ 16. Compared to previous works using other large X-ray surveys, such as the Cosmic Evolution Survey (COSMOS), we only select sources as CT candidates when over 50\% of their \nh\ PDF is above $1.5\times10^{24}\; \rm cm^{-2}$. Our criterion is much stricter than the one adopted in \citet{lanz18ct}, which selected 67 CT AGN candidates and estimated a total number of $\approx$ 38 CT sources from COSMOS with at least 5\% of their \nh\ PDF above \nh\ $= 10^{24}\; \rm cm^{-2}$.

Furthermore, to quantitatively examine the sensitivity of our results to the adopted priors in the Bayesian analysis, we measure the difference between the prior and the posterior distributions. A useful metric for this purpose is the Kullback-Leibler divergence ($D_{\rm KL}$), defined as $D_{\rm KL}=$ \(\int_{-\infty}^{+\infty} pr(x) \ln\frac{pr(x)}{po(x)} \,dx\) for prior $pr(x)$ and posterior $po(x)$ (e.g., \citealt{leja19}). With higher $D_{\rm KL}$, the posterior and the prior become increasingly divergent. We calculate $D_{\rm KL}$ for our CT candidates using the flat \nh\ distribution prior (Section 2.1), resulting in a value of 2.3 (much larger than the minimum value $\approx$ 0.05), which suggests that the posterior \nh\ distribution is not significantly similar to its prior. We also test the dependence of the posterior \nh\ outputs on priors by adopting a Gaussian prior. In this case, we obtain a $D_{\rm KL}$ of 2.1 and still successfully select the 22 CT candidates. Therefore, our results are not heavily dependent upon the flat \nh\ prior chosen for our analyses.

Moreover, we examine the quality of the photometric redshifts of our selected CT and HO samples, when spectroscopic redshifts are not available. Most of the selected sources show small uncertainties with $peak-prob\approx1$ and $Q_z<0.2$, suggesting high-quality photometric redshifts with single peaks. For a few HO sources with smaller $peak-prob$ and higher $Q_z$, we further examine the X-ray spectrum (e.g., Fe K$\alpha$ emission at 6.4 keV) to ensure that the spectral fits support the adopted redshift. Therefore, the posterior \nh\ PDFs do not appear heavily affected by the uncertainties of photometric redshifts.

\subsection{Additional observations with XMM-Newton}

Thanks to additional observations by XMM-Newton, we obtained a 42-hour exposure starting from 2022/05/16 23:33 on a selected target J003836.99-433709.8 (XID: ES1759) in the ELAIS-S1 field. The target was selected based on our preliminary spectral analyses in the ELAIS-S1 field as an attractive heavily-obscured AGN candidate (\nh\ $\approx 10^{23.7}\; \rm cm^{-2}$) with accurate and relatively high redshift. This source has a photometric redshift of $2.33^{+0.27}_{-0.17}$ and is relatively bright in the \textit{R}-band. With the additional exposure time, we obtain a total full-band (0.5--10 keV) counts of 1133, including the previous observations in the ELAIS-S1 field.

\begin{figure*}[t]
\epsscale{1.1}
\plotone{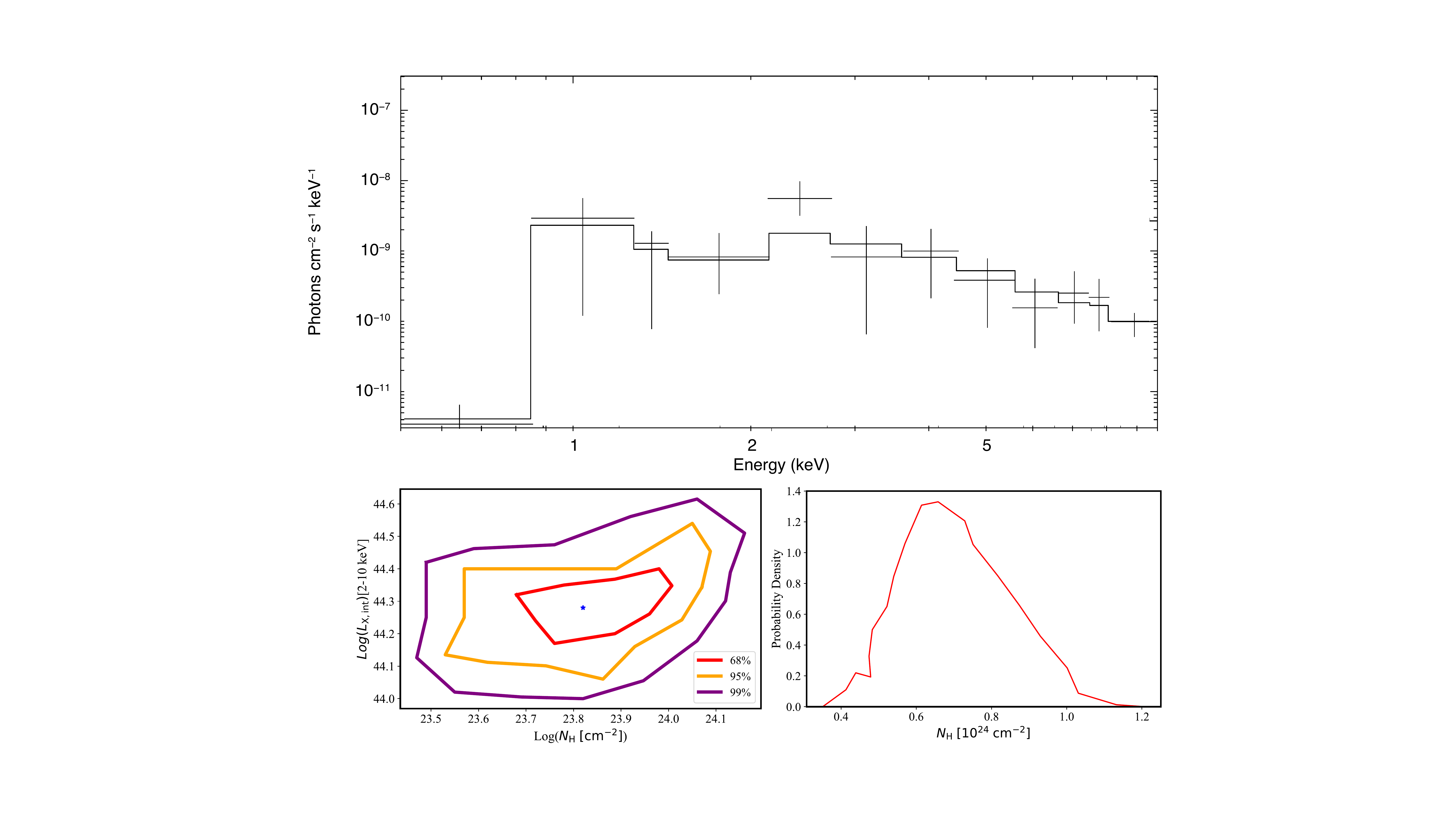}
\caption{The analysis results for J003836.99-433709.8 at $z=2.333$, a candidate with heavy obscuration selected by X-ray spectral analysis. \textit{Upper}: Unfolded combined X-ray spectrum of MOS1 and MOS2 with the best fit (binned only for visualization). \textit{Lower Left}: The Bayesian analyses result for the same candidate with a best-fit \nh\ of $10^{23.83}\; \rm cm^{-2}$. \textit{Lower Right}: The probability distribution function of \nh\ obtained from BXA, with over 50\% of the \nh\ PDF above $5\times10^{23}\; \rm cm^{-2}$. Thus, the \nh\ probability density confirms that this source is heavily-obscured. \label{fig:example}}
\end{figure*}

The additional observation significantly enhances our target's total counts by a factor of $\approx$ 3 and provides an increased SNR, which is essential for deriving a solid estimate of \nh\ through spectral fitting. With this recent observation, the derived probability distribution of \nh\ for this target is much better constrained. As a result, the best-fit $\Gamma$ from a simple power-law fit is $-0.5$, confirming the target has one of the hardest X-ray spectra in the ELAIS-S1 field. Based on the Bayesian analyses described in Section 2.1, we derive its posterior probability distribution and the PDF with a best-fit \nh\ value of $10^{23.83}\; \rm cm^{-2}$ (shown in Figure~\ref{fig:example}), indicating this AGN is indeed heavily obscured. The consistency between the improved analyses and our preliminary estimate of \nh\ confirms the effectiveness of our selection method described in the previous subsection. We further estimate the X-ray redshift from the spectrum to examine the accuracy of the photometric redshift. We follow the strategy in \citet{peca21} and include a redshifted Gaussian component (zguass) for the Fe K$\alpha$ emission line at 6.4 keV in the rest frame. The line width is fixed at $\sigma = 10$ eV to only consider the narrow component since the broad emission component should be absorbed due to the heavy obscuration. We then let the redshift vary and obtain the X-ray redshift by a combination of signatures due to absorption, such as the Fe K$\alpha$ emission and the absorption edge, in case of heavy obscuration. The best-fit X-ray redshift is 2.15, which is in agreement with the photometric redshift.

\begin{deluxetable*}{lcccc}
\tabletypesize{\scriptsize}
\tablecaption{Numbers of selected AGNs in each field of \servs\ }
\tablewidth{0pt}
\tablehead{
\colhead{Field} & \colhead{CT candidates} & \colhead{CT candidates with spec-z} &\colhead{Heavily-obscured candidates} &\colhead{Heavily-obscured candidates with spec-z}}
\startdata
XMM-LSS & 12 & 4  & 58 & 19\\
W-CDF-S &7 & 3  & 43 & 15\\
ELAIS-S1 & 3 & 1 & 35 & 10
\\[0.01mm]
\tableline
\\[0.01mm]
Total & 22 & 8  & 136 & 44
\enddata 
\label{table:num}
\end{deluxetable*}

\begin{deluxetable*}{llcccccccccc}
\tabletypesize{\scriptsize}
\tablecaption{Heavily-obscured AGNs in each field of \servs\ }
\tablewidth{0pt}
\tablehead{
\colhead{ObjectID} & \colhead{Field}& \colhead{RA}& \colhead{DEC} & \colhead{$z$}& \colhead{Net Counts} &\colhead{$\Gamma$ \tablenotemark{a}} &\colhead{$\log{L_X}$ \tablenotemark{b}} &\colhead{$\log{L_X,cor}$ \tablenotemark{c}}&\colhead{$\log{N_{\rm H}}$}  & \colhead{$P_{HO}$} & \colhead{$\log{L_{6\mu {\rm m}}}$} \\
\colhead{} & \colhead{} & \colhead{} &\colhead{} &\colhead{} & \colhead{} & \colhead{} & \colhead{$({\rm erg}\; {\rm s}^{-1})$} & \colhead{$({\rm erg}\; {\rm s}^{-1})$}&\colhead{$\rm (cm^{-2})$} &  \colhead{\%}  & \colhead{$({\rm erg}\; {\rm s}^{-1})$} \\}
\startdata
XMM03252 & XMM-LSS & 35.9545 & -5.19429 &$0.43^{+0.03}_{-0.02}$  & 100  & -0.3 & 42.5 & 43.3  & $23.57^{+0.45}_{-0.17}$  & 81 & 43.6\\
WCDFS2220 & W-CDF-S & 53.1299 & -27.2823 & $0.60^{+0.07}_{-0.08}$& 132  & -0.3 & 42.4 & 43.1& $24.03^{+0.22}_{-0.20}$ & 97  & 43.3\\
ES0211 & ELAIS-S1 & 9.03463 & -44.3403 &$0.47^{+0.04}_{-0.08}$  & 156  & -0.2 &  42.5 & 43.5 & $23.65^{+0.77}_{-0.15}$ & 55& 43.8  \\
\enddata

\tablenotetext{a}{$\Gamma$: best-fit value using a simple power-law model}
\tablenotetext{b}{$\log{L_X,obs}$: observed X-ray luminosity in the rest frame without absorption correction}
\tablenotetext{c}{$\log{L_X,cor}$: intrinsic X-ray luminosity, corrected using the best-fitted absorption}

\tablecomments{The full table contains 15 columns of information for the 136 heavily-obscured AGNs, including redshift flag and $R$-band magnitude. Only a portion of this table (one example for each field) is shown here to demonstrate its form and content. A machine-readable version of the full table is available.}

\end{deluxetable*}

\begin{deluxetable*}{llcccccccccc}
\tabletypesize{\scriptsize}
\tablecaption{Selected CT AGN candidates in each field of \servs }
\tablewidth{0pt}
\tablehead{
\colhead{ObjectID} & \colhead{Field}& \colhead{RA}& \colhead{DEC} & \colhead{$z$}& \colhead{Net Counts} &\colhead{$\Gamma$ \tablenotemark{a}} &\colhead{$\log{L_X}$ \tablenotemark{b}} &\colhead{$\log{L_X,cor}$ \tablenotemark{c}}&\colhead{$\log{N_{\rm H}}$}  & \colhead{$P_{\rm CT}$} & \colhead{$\log{L_{6\mu {\rm m}}}$} \\
\colhead{} & \colhead{} & \colhead{} &\colhead{} &\colhead{} & \colhead{} & \colhead{} & \colhead{$({\rm erg}\; {\rm s}^{-1})$} & \colhead{$({\rm erg}\; {\rm s}^{-1})$}&\colhead{$\rm (cm^{-2})$} &  \colhead{\%}  & \colhead{$({\rm erg}\; {\rm s}^{-1})$} \\}
\startdata
XMM01845  & XMM-LSS & 35.1974 & -4.37777   &    $1.26^{+0.01}_{-0.01}$   & 217 & -0.5 &  42.4 & 43.8  & $24.14^{+0.54}_{-0.13}$   & 57 & 44.4\\
WCDFS1629 & W-CDF-S & 53.3614 & -28.5692 &  $1.01^{+0.03}_{-0.06}$ & 448  & -0.6 &  41.8 & 43.2  &$24.24^{+0.37}_{-0.11}$    & 65 & 43.4 \\
ES1278 & ELAIS-S1 & 53.3614 & -28.5692 &   $0.73^{+0.09}_{-0.03}$  & 325  & -0.6 &  42.7 & 44.2 &$24.35^{+0.56}_{-0.22}$     & 78& 44.9  \\
\enddata

\tablecomments{The full table contains 15 columns of information for the 22 CT AGNs. Only a portion of this table (one example for each field) is shown here to demonstrate its form and content. A machine-readable version of the full table is available.}

\tablenotetext{a}{$\Gamma$: best-fit value using a simple power-law model}
\tablenotetext{b}{$\log{L_X,obs}$: observed X-ray luminosity in the rest frame without absorption correction}
\tablenotetext{c}{$\log{L_X,cor}$: intrinsic X-ray luminosity, corrected using the best-fitted absorption}

\label{table:eg}

\end{deluxetable*}


\section{Multiwavelength SED fitting}

\begin{figure*}[t]
\epsscale{1.1}
\plotone{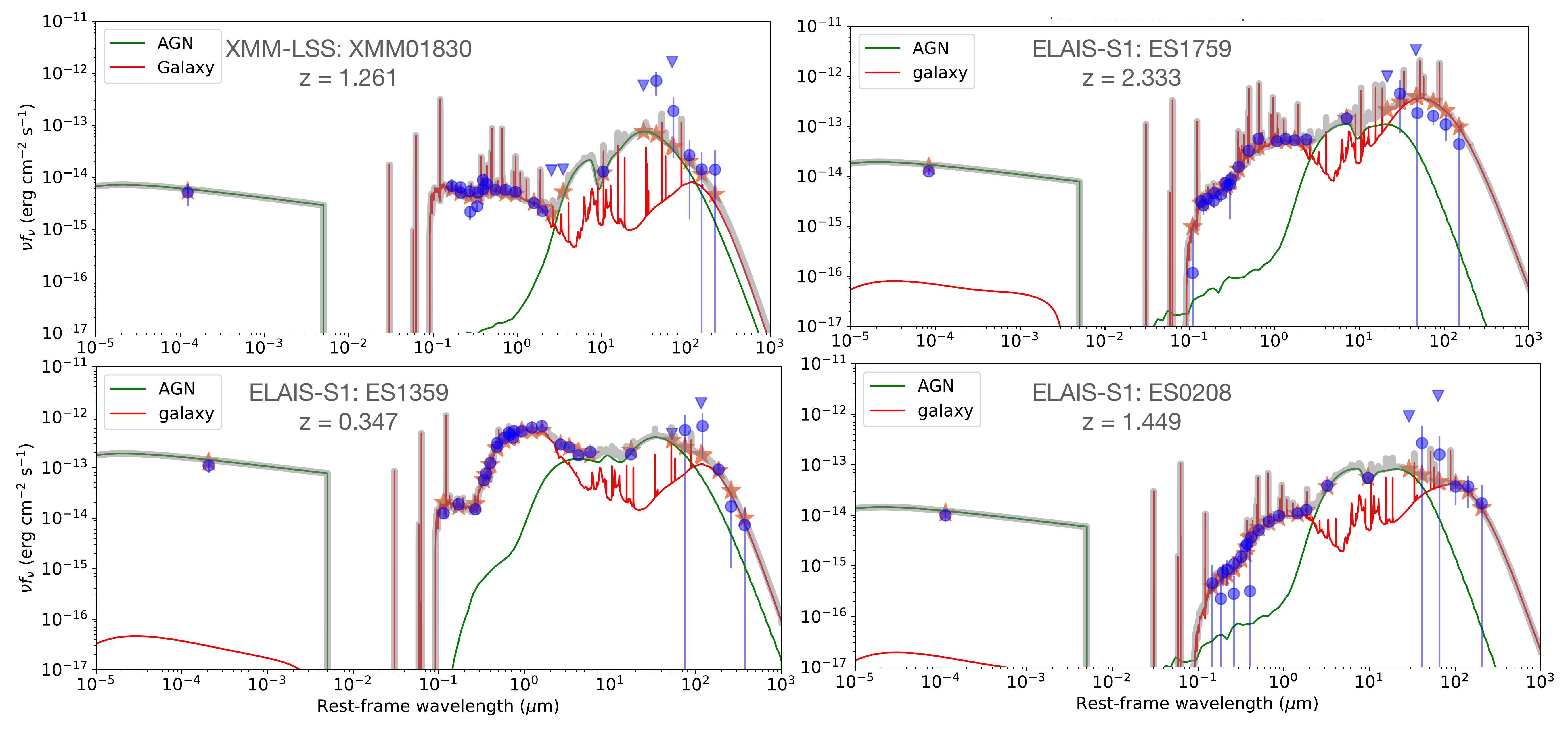}
\caption{Examples of the SED fitting of our selected CT (\text{right}) and heavily-obscured AGNs (\text{left}) with default parameters for the general AGN population (details in \citealt{zou22sed}). The blue points are the observed photometry and the triangles are upper limits. The thick grey line represents the best-fit model, including an AGN component shown as the green line and a galaxy component shown as the red line. As indicated in the figures, the extinction of the AGN component in the UV-to-optical bands suggests the existence of heavy absorption, which is consistent with our results from the X-ray spectral analyses. \label{fig:sed}}
\end{figure*}

\begin{figure*}[t]
\epsscale{1.1}
\plotone{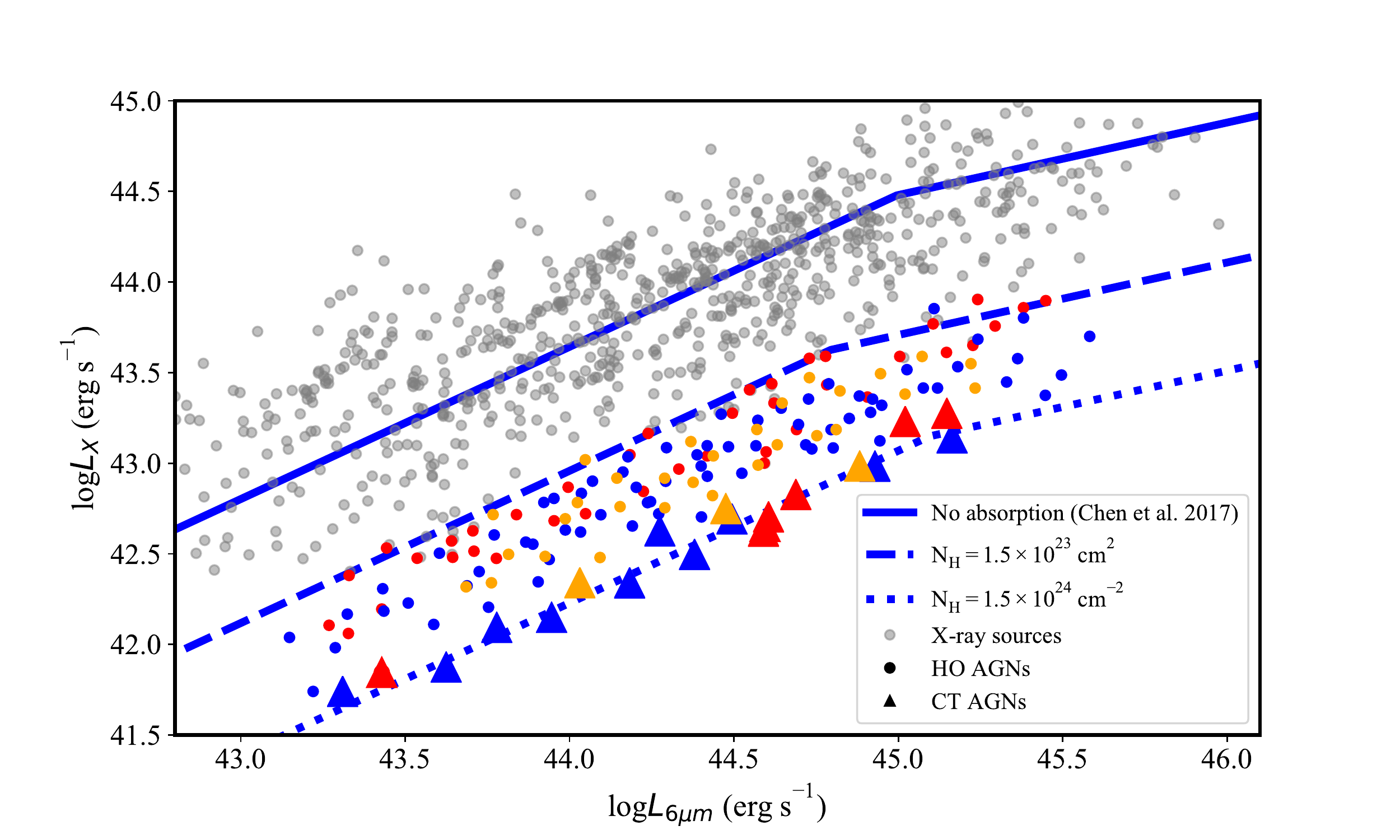}
\caption{Observed X-ray luminosity vs IR luminosity in three \servs\ fields. Triangles and circles are the CT AGN candidates and heavily-obscured sample, respectively, with multi-wavelength coverage. The blue, red and orange colors mark the selected CT candidates in the XMM-LSS, W-CDF-S, and ELAIS-S1 field, respectively. We also show the rest of detected AGNs (grey dots) in W-CDF-S for comparison. The blue solid line is taken from \citet{chen17} with no absorption. The blue dashed line is heavily-obscured relation derived with the \textit{borus} model with a column density of $1.5\times10^{23}\; \rm cm^{-2}$, and the dotted line is CT relation with a column density of $1.5\times10^{24}\; \rm cm^{-2}$. The low observed hard X-ray luminosities of CT AGNs imply very heavy obscuration in all candidates, which is consistent with our results from the X-ray spectral analyses. \label{fig:lxlir}}
\end{figure*}

All fields in the XMM-SERVS survey are well-covered with multi-wavelength surveys (many listed in Table 1 of \citealt{zou22sed}). We use \textit{CIGALE} (e.g., \citealt{yang22sed}) for SED fitting, and one of the significant advantages of this approach is that it allows inclusion of X-ray photometry from \servs . More details about the SED fitting process have been published in \citet{zou22sed}, which aims to provide high-quality estimates of AGN and host-galaxy characteristics for the general AGN population, including both Type 1 and Type 2 AGNs. 

Examples of the SED best fits for our heavily-obscured sample and CT candidates are shown in Figure~\ref{fig:sed}. These preliminary SED fitting results (Figure~\ref{fig:sed}) are derived with default fitting parameters \citep{zou22sed}. The best-fit SED AGN inclination angles of our CT and HO samples are above 50 degree, suggesting Type 2 AGNs. Furthermore, for 80\% of our CT candidates and over 50\% of the HO sample, the spectral shape in the UV-to-optical bands (shown in Figure~\ref{fig:sed}) suggests heavy absorption in general, which is mainly from the torus component, consistent with our X-ray spectral analysis results. 

In order to reflect accurately the observed heavy obscuration and its effects upon the multi-wavelength photometry (such as the derived intrinsic mid-IR luminosity), we adjust the absorption parameters in the SED model specifically for these CT and HO AGNs. First, we derive the X-ray photometry by calculating the absorption-corrected X-ray flux at rest-frame 2 keV, which is used in the SED fitting with \textit{CIGALE}. We then made modifications of the SED fitting by adopting more sophisticated galaxy IR templates \citep{drai14}, limiting the AGN inclination to Type 2 views only, and allowing the AGN torus optical depth to vary. Finally, from our systematic fits to the well-sampled SEDs of all AGNs in \servs , we obtain the rest-frame 6 $\rm \mu$m luminosities ($L_{6 \mu \rm m}$) and host properties (stellar mass, $M_*$, and star formation rate, SFR) of our candidates. Note that we only perform SED fitting for the sources with excellent multi-wavelength coverage. 2 out of 22 CT candidates and 14 out of 136 heavily-obscured sources do not have SED fits and derived characteristics (e.g., $L_{6 \mu \rm m}$) due to the lack of VISTA Deep Extragalactic Observation Survey (VIDEO) coverage, which is necessary for obtaining quality forced photometry and sufficient SED coverage in wavelength \citep{zou22sed}.


Based on established relationships between intrinsic mid-IR and observed \mbox{X-ray} luminosities in the rest frame (e.g., \citealt{chen17}), we derive the relation between the observed \mbox{X-ray} luminosity $L_X$ in the rest frame and $L_{6 \mu \rm m}$ with different levels of obscuration. To compute the expected suppression of the hard X-ray luminosity as a function of \nh , we consistently adopt \textit{Borus02} to derive the luminosity relations for \nh\ $=\; 1.5\times10^{24}\; \rm cm^{-2}$. For all AGNs detected in \servs , including the selected HO and CT candidates, we calculate their \mbox{X-ray} luminosity in the rest frame based on the best-fit $\Gamma$ obtained from our spectral fits in Section 2. As a result, the $L_X$ values of all our CT sample show heavy suppression compared to their $L_{6 \mu \rm m}$, indicating the existence of heavy obscuration (\citealt{alex08ct, lans17nustar}). All HO and CT candidates locate close to the expected suppressed relation at the heavily-obscured level and the CT level ($N_{\rm H} = 1.5\times10^{24} \; \rm cm^{-2}$; Figure~\ref{fig:lxlir}), respectively, consistent with our findings from X-ray spectral analyses (Section 2). We also notice that after absorption corrections, the HO and CT candidates show high intrinsic X-ray luminosity, which suggests that sources with intrinsically low X-ray luminosity and extreme obscuration are still missing or undetected in the current observations.

\section{Discussion}

\subsection{Intrinsic column density distribution}

After identifying sources with heavy obscuration from X-ray observations, we review the intrinsic distributions of these selected samples compared to the general AGN population. We first derive the probability distribution of the count rate for each source, $P(CR)$, using Equation~5 in \citet{vito18fct}. Then we assess the probability of source detection by the X-ray detector in each field of XMM-SERVS at given full-band flux ($f_{FB}$), which is defined as $P_{det}(f_{FB})$ with the following equation

\begin{equation}
P_{det}(f_{FB}) = \frac{1}{2}(\rm erf{(b * ({\log f_{FB}}-a}))+1)
\end{equation}
where $a$ and $b$ are free parameters to be measured using sources with full-band detections in the XMM-SERVS fields (Zou et al. submitted). 

To eliminate vignetting effects in the outskirts of the fields, which affect the quality of the observational data, we only consider detected sources with the summation of the exposure from PN, MOS1, and MOS2 to be more than 45 ks in each field. The inner regions of the three fields have better multi-wavelength data and identification accuracy with VIDEO coverage. The two parameters, $a$ and $b$, are $(-14.42, 4.32)$, $(-14.35, 4.69)$, and $(-14.30, 3.72)$ for XMM-LSS, W-CDF-S, and ELAIS-S1, respectively. 

Unlike $f_{FB}$, which is sensitive to the assumed spectral shape, count rate is a more fundamental measurement from X-ray observations obtained directly from images. Therefore, we obtain the probability of a source with certain count rate being detected, $P_{det}(CR)$,  using

\begin{equation}
P_{det}(CR)= \frac{P_{det}(f_{FB})}{f_X}
\end{equation}
where $f_X$ is the conversion factor from count rate to flux for general sources with a nominal distribution.

Although the detection probability can also be derived by estimating the fraction of the total survey area with sensitivities deeper than a given flux based on the sensitivity curve, the obtained sensitivity is biased by the chosen aperture size and the complications of real source detection.

We use the probability of count rate $P^i(CR)$ of each source $i$ to weight the source detection, in order to consider the different sensitivity in different fields of a survey. Following the approach in \citet{vito18fct}, we define $P^i(CR)$ as :

\begin{equation}
P^i(CR)= \int \frac{P(CR)}{P^i_{det}(CR)}dCR
\end{equation}

In each obscuration bin, we then derive the intrinsic distribution of \nh\ in each field using 

\begin{equation}
N(\log N_{\rm H}) = \sum \frac{P^i(\log N_{\rm H})}{P^i(CR)}
\end{equation}
where $P^i(\log N_{\rm H})$ is the probability distribution of \nh\ for each source $i$, obtained from our spectral analysis in the previous section.

\begin{figure}
\includegraphics[scale=0.5]{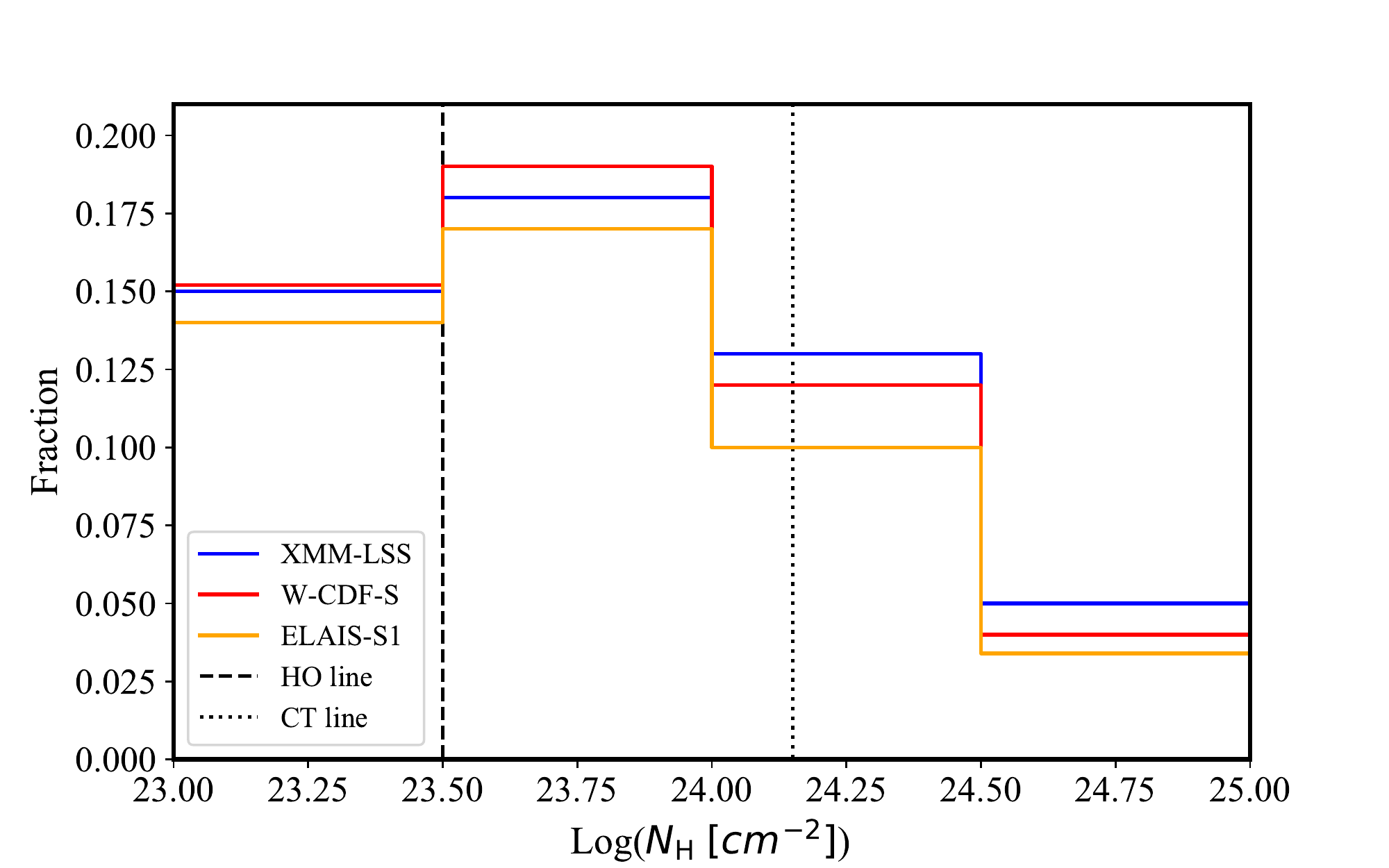}
\caption{Intrinsic distribution of column density for our sample of detected obscured AGNs. The dashed line represents the threshold to select HO AGNs, and the dotted line represents the threshold for CT AGNs used in this work. Blue, red, and orange colors show the fraction of sources in each \nh\ bin in the three fields (XMM-LSS, W-CDF-S, and ELAIS-S1), respectively. Within uncertainties of 68\% confidence level derived by bootstrap, the \nh\ distributions of the three fields covered in the XMM-SERVS are consistent. \label{fig:nh}}
\end{figure}

We use the bootstrap to compute the corresponding errors with 1000 iterations in each \nh\ bin, and select the 16th and 84th percentile values as the 68\% confidence interval of $N(\log N_{\rm H})$. The intrinsic \nh\ distribution of obscured AGNs is shown as fractions in Figure~\ref{fig:nh}.

\subsection{The fraction and space density of CT sources}

Following the approach in \citet{vito18fct}, we derive the space density for the full XMM-SERVS sample in the CT regime, which is the number of CT sources, $N_{CT}(\log N_{\rm H})$, integrated by the following equation in the specific redshift, \nh , and luminosity parameter space divided by the volume sampled by the survey at the particular redshift bin: 

\begin{equation}
\frac{N_{CT}(\log N_{\rm H})}{V} = \int_{\log N_{\rm H} = 24}^{25} \frac{P^i(L_{X}, \log N_{\rm H})}{P^i_{det}(CR)} d\log N_{\rm H}/V
\end{equation}
where $L_X = \frac{f_X}{l_X}$CR is the intrinsic X-ray luminosity, and $l_X$ is the conversion factor between flux and luminosity. Like $f_X$ used in equation (2), the value of $l_X$ is obtained for each source from spectral fitting with XSPEC, determined by \nh , redshift, and photon index $\Gamma$. We use a similar bootstrap approach as described in the previous subsection to estimate the 90\% confidence intervals for comparison with previous research.

\begin{figure}
\includegraphics[scale=0.2]{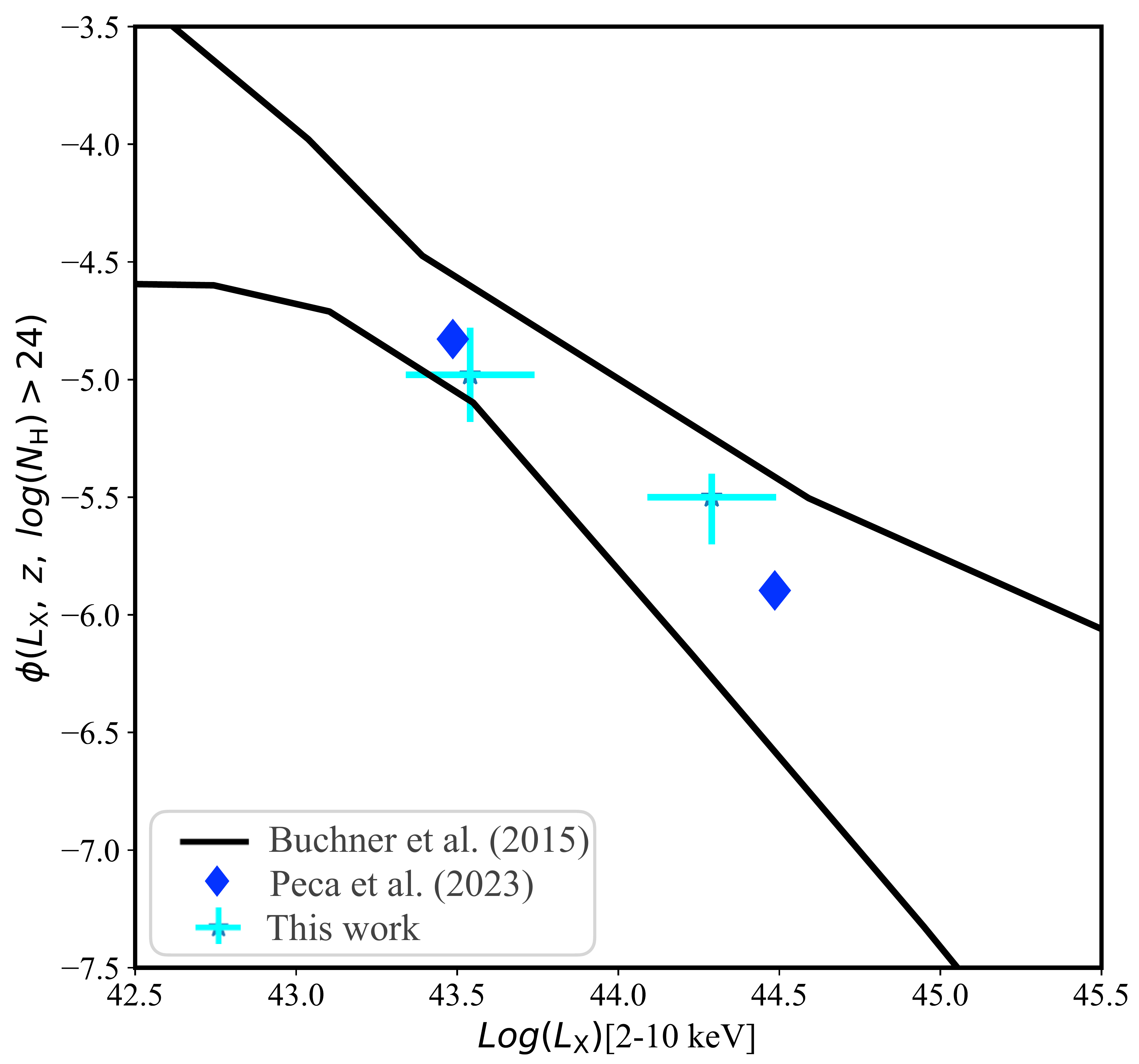}
\caption{Space density of AGN per luminosity bin in the CT regime at $z<1.5$. The cyan points show results from this work in two luminosity bins. The median redshift of the sample sources is 0.7. The error bars of the space density are adopted as the 90th percentiles derived by bootstrap around the median of the space density probability distribution function. The black lines show the CT X-ray luminosity function at $z =$ 0.5--0.75 from \citet{buch15} with 90\% credible intervals for comparison, and the blue diamonds show the luminosity function from \citet{peca22xlf}. Our results are in agreement with these works.
\label{fig:fct}}
\end{figure}

To focus on sources with heavy obscuration, we only consider the high-\nh\ regime ($10^{24}$ \cdens\ $<$ \nh\ $<\;10^{25}$ \cdens) in our space-density calculations up to $z=1.5$. After dividing the integrated number of sources by the volume, Figure~\ref{fig:fct} shows the AGN space density as a function of 2--10 keV $L_X$ up to $z=1.5$ for the CT regime ($10^{24}\; \rm cm^{-2}$ $<$ \nh\ $<\;10^{25}\; \rm cm^{-2}$). Our estimate is overall in agreement with the work of \citet{buch15} as well as the space density of the obscured AGN population binned by redshift and X-ray luminosity in \citet{peca22xlf}. When we divide the redshift range $z<1.5$ into two smaller redshift bins ($z>0.75$ and $z<0.75$), the space density shows an increase with redshift. We also notice that toward higher redshifts, the space density constraints tend to suffer from large uncertainties due to the small number of counts and large photometric redshift errors, especially for CT sources. To further improve and extend the estimates for heavily-obscured AGNs at higher redshift, follow-up spectroscopic observations and future deeper photometry (e.g., from LSST, Euclid, and Roman observations) are essential to obtain better estimates of redshift and constraints on the AGN space density, especially at higher redshift (e.g., $z>2$).

We also derive the AGN CT fraction in each field as $f_{CT}(z, L_{X}) = \frac{N_{CT}}{N_{tot}}$, where $N_{CT}$ is the estimated number of CT sources calculated from equation 4 in the field with intrinsic $\log N_{\rm H} > 24$, and $N_{tot}$ is the total number of detected sources in the field integrated from $\log N_{\rm H} = 20$ to $\log N_{\rm H} = 26$. We note that sources with log(\nh )$>$ 25 at $z>1$ are not complete in the observations, resulting in large error bars. However, thanks to our representative AGN population, $f_{CT}$ derived up to log(\nh )= 25 is a good approximation for the full CT sample. We list our results in Table 4, which shows that $f_{CT}$ is consistent in the three fields, increasing with redshift while decreasing with X-ray luminosity. Our results are in agreement with previous works, such as \citet{lalo23}, which obtained an increasing $f_{CT}$ from 0.21 to 0.40 at redshift $z<0.5$ and higher redshift (up to $z=2.5$), respectively.

\begin{deluxetable}{lccccccc}
\tabletypesize{\scriptsize}
\tablecaption{CT fraction in different redshift and intrinsic luminosity bins.}
\tablewidth{0pt}
\tablehead{
\colhead{Bin} & \colhead{$z_{median}$} & \colhead{$\log L_{X,median}$} &\colhead{XMM-LSS}&  \colhead{W-CDF-S} &\colhead{ELAIS-S1} &\colhead{Total}\\
\colhead{} & \colhead{} & \colhead{} & \colhead{} & \colhead{} & \colhead{} & \colhead{}  \\
}
\startdata
$z\leq0.75$  & - & 43.6 & $0.27^{+0.03}_{-0.02}$ & $0.28^{+0.03}_{-0.02}$  & $0.22^{+0.04}_{-0.03}$ & $0.26^{+0.02}_{-0.02}$\\
$z>0.75$ &  - & 44.2 & $0.47^{+0.03}_{-0.04}$ & $0.48^{+0.03}_{-0.03}$ & $0.44^{+0.05}_{-0.04}$& $0.44^{+0.02}_{-0.03}$\\
$\log L_{X} \leq 44$   & 0.7  & - & $0.38^{+0.02}_{-0.04}$ & $0.35^{+0.05}_{-0.04}$   & $0.34^{+0.04}_{-0.04}$ & $0.36^{+0.02}_{-0.03}$\\
$\log L_{X}>44$  &  1.1 & -  & $0.18^{+0.03}_{-0.04}$ & $0.22^{+0.04}_{-0.05}$ & $0.17^{+0.03}_{-0.03}$ & $0.24 ^{+0.02}_{-0.02}$
\enddata 

\label{table:fct}
\end{deluxetable}

\begin{deluxetable*}{lcc}
\tabletypesize{\scriptsize}
\tablecaption{Host-galaxy properties at different obscuration levels.}
\tablewidth{0pt}
\tablehead{
\colhead{Obscuration Level} & \colhead{$\log (SFR)_{median}$}&  \colhead{$\log (M_*)_{median}$} \\
\colhead{} & \colhead{$M_\sun/year$} & \colhead{$M_\sun$} \\
}

\startdata
Less obscured  & $0.26_{-0.04}^{+0.07}$ &$10.73_{-0.06}^{+0.10}$\\
HO & $0.23_{-0.03}^{+0.04}$ & $10.85_{-0.05}^{+0.09}$ \\
CT & $0.28_{-0.02}^{+0.04}$ & $10.95^{+0.10}_{-0.08}$ \\
\enddata 

\label{table:gal}
\end{deluxetable*}

\subsection{Galactic properties}

We examine if our CT AGNs have different host-galaxy $M_*$ and SFR compared with less-obscured X-ray AGNs in this section. Since their $z$ and $L_{X, int}$ distributions are different, we need to control for these parameters to avoid possible differences in host-galaxy properties caused by different $z$ and $L_{X, int}$ distributions (e.g., \citealt{zou19}). Since the size of our CT candidate sample is much smaller than the total number of XMM-SERVS AGNs, we can locally construct a sufficiently large comparison sample for each of our CT sources by selecting HO and less-obscured XMM-SERVS AGNs within the corresponding $z-\log L_{X, int}$ bin, where the bin sizes are $\delta z = 0.01$ and $\delta \log L_{X, int} = 0.05$. We first obtain a random bootstrap sample from our 22 CT AGNs and randomly select 5 comparison sources from the HO and less-obscured samples, respectively, allowing duplicates for each CT AGN. The resulting HO and less-obscured sample (each with size $22*5$) is our comparison sample in one random realization. Meanwhile, we obtain another bootstrap sample of our 22 CT AGNs and compare their median $M_*$ and SFR with those of the corresponding comparison sample. We repeat this procedure 5000 times. In these 5000 iterations, we find 4670 times and 4209 times that CT and HO sources, respectively, have a larger median value of $M_*$ than the less-obscured sample. We list the median $M_*$ and SFR values of the controlled sample in Table 5.

Therefore, we find that $M_*$ appears likely to increase with absorption level. When comparing CT and less-obscured sources, the median values of $M_*$ are 93.4\% tentatively different but cannot be statistically confirmed due to a limited sample size. We do not see such a difference for SFR in the three samples with different obscuration. 

\subsection{Soft excess}

We also note that over 90 sources in our sample show significant evidence for a soft X-ray excess component, which requires a second power-law to fit the excess. More than half of these sources are heavily obscured. However, we do not find a significant relation between the soft excess and \nh . The soft excess is believed to originate often from scattering by the obscuring and other circumnuclear material. To confirm that the missing correlation is not caused by large uncertainties and double peaks of the \nh\ value due to flux limits and low SNR, we perform simulations of 100 fake spectra for each source with the same redshift and \nh\ with 50, 150, and 200 counts between 0.5--2 keV. We then fit the spectra and derive the best-fit \nh\ using the same procedure described in the previous section. We still find no significant correlation between the soft excess component and \nh . However, we find a weak correlation between opening angles and the soft excess component. Sources with small opening angles tend to show more soft excess, which indicates a possible relation between scattering from obscuring materials and torus geometry (e.g., \citealt{brig12}).

\section{Conclusion}

In this work, we extract X-ray spectra and perform detailed spectral analyses of all sources in the three \servs\ fields in a uniform and systematic way. We adopt Bayesian analyses with the physical torus model \textit{Borus02} to derive AGN characteristics, in order to select CT candidates with \nh  $\;>\;1.5\times10^{24}$ \cdens\ as well as a representative sample of heavily-obscured AGNs.  As a result, we uncover 22 CT AGN candidates with good signal-to-noise ratios as well as a large representative sample of heavily-obscured AGNs. Most of our CT candidates and over half of the HO sample show dust obscuration in their SED fits with default parameters, which with consistent to their X-ray spectral properties. We further obtain an increasing \fct\ from low to high redshift. By continuing studies of the selected obscured samples in the LSST Deep-Drilling Fields with future observations, it will shed light on the connection between AGN obscuration and host galaxy evolution.

\acknowledgments

We thank the anonymous referee for constructive suggestions and comments. WY, WNB, and FZ acknowledge financial support from NASA grant 80NSSC19K0961, NSF grant AST-2106990, and CXC grant AR1-22006X. B.L. acknowledges financial support from the National Natural Science Foundation of China grant 11991053. FEB acknowledges support from ANID-Chile BASAL CATA FB210003, FONDECYT Regular 1200495 and 1190818, and Millennium Science Initiative Program  – ICN12\_009 . RCH acknowledges support from the National science Foundation through CAREER Award number 1554584.

\bibliographystyle{apj}

 \newcommand{\noop}[1]{}

\end{document}